\documentclass[prd,aps,twocolumn,nofootnibib]{revtex4}
\usepackage{graphicx}
\usepackage{url}
\usepackage{color}
\def\epm{$e^+e^-$}

\def\mwcda{$\mu$WCDA}

\begin{document}
\title{Possibility of measurement of cosmic ray electron spectrum up to 100 TeV with two-layer water Cherenkov detector array }
\author{Andrii Neronov$^{1,2}$,  and Dmitri Semikoz$^1$ }
\affiliation{$^1$Université de Paris, CNRS, Astroparticule et Cosmologie,  F-75006 Paris, France\\
$^2$Astronomy Department, University of Geneva, Ch. d'Ecogia 16, 1290, Versoix, Switzerland}

\begin{abstract}
Measurements of cosmic ray electron$+$positron spectrum above  10 TeV with ground-based experiments is challenging  because of the difficulty of rejection of hadronic extensive air shower background. We study the efficiency of rejection of the hadronic background with water Cherenkov detector array supplemented by muon detection layer. We show that addition of a  continuous muon detection layer to the experimental setup allows to achieve a $\sim 10^{-5}$ rejection factor for hadronic background at 10~TeV and enables  measurement of electron spectrum in 10-100~TeV energy range. We show that measurements of electron spectrum in this energy range do not require a high-altitude experiment and can be done with a sea-level detector.     
\end{abstract}
\maketitle

\section{Introduction}

Measurements of cosmic ray electron$+$positron spectrum using space and ground-based detectors are currently available up to 10~TeV energy range (see Ref. \cite{Lipari:2019abu} for a review). The space-based detectors, like Fermi/LAT \cite{Abdollahi:2017nat}, AMS-02 \cite{Aguilar:2019ksn}, CALET \cite{Adriani:2017efm}, DAMPE \cite{Ambrosi:2017wek}, run out of signal statistics already at TeV energies. At higher energy, ground-based measurements have been done with HESS \cite{Aharonian:2008aa} and VERITAS \cite{Archer:2018chh} Imaging Atmospheric Cherenkov Telescope (IACT) systems up to approximately 5~TeV, and possibly to higher, $\sim 20$~TeV, energy by HESS \cite{kerszberg_icrc}.

\begin{figure}
\includegraphics[width=\columnwidth]{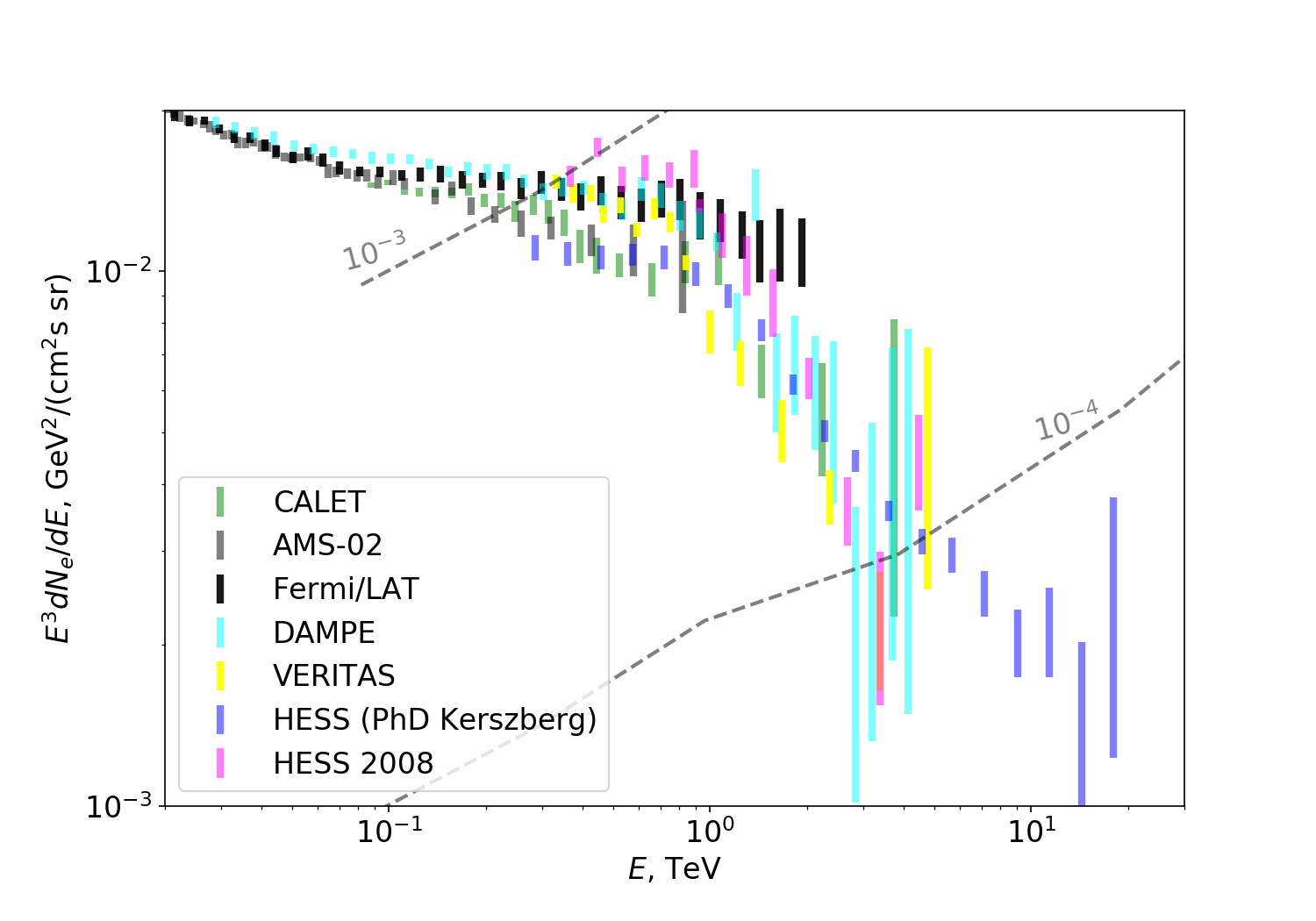}
\caption{Measurements of \epm\ spectrum by AMS-02  \cite{Aguilar:2019ksn}, Fermi/LAT \cite{Abdollahi:2017nat}, DAMPE  \cite{Ambrosi:2017wek}, CALET \cite{Adriani:2017efm}, HESS \cite{Aharonian:2008aa},  \cite{kerszberg_icrc}, VERITAS \cite{Archer:2018chh}.  Dashed lines show different levels of rejection of backgorund of cosmic ray nuclei needed to detection of flux as a given level.  }
\label{fig:electron_spectrum}
\end{figure}

These measurements reveal  spectral features in the TeV band (Fig. \ref{fig:electron_spectrum}),  in part formed by an interplay between synchrotron and inverse Compton cooling of electrons  in the interstellar magnetic and radiation fields \cite{1985Afz....23..479A,2010ApJ...710..236S}, but perhaps also by further effects of acceleration of primary electrons or production of secondary electrons and positrons in yet uncertain class of astronomical sources and their subsequent  propagation through the Galaxy. The positron flux measured by AMS-02  is  much lower than the overall electron plus positron flux level \cite{Aguilar:2019ksn}, thus the TeV softening of the spectrum visible in Fig. \ref{fig:electron_spectrum} certainly carries information about the nature of the electron source(s).  The TeV suppression can be a feature introduced by the discreetness and intermittent nature of distribution of electron  sources in the local Galaxy, like pulsars and supernova remnants \cite{1995PhRvD..52.3265A} or can also be an average high-energy cutoff in the intrinsic spectra the source(s). It also can be an isolated feature superimposed on otherwise powerlaw-like spectrum. e.g. by a contribution from a specific nearby source \cite{Recchia:2018jun}.  

To distinguish between these different possibilities, a better characterisation of the spectrum and reliable measurements of the flux extending well into the 10-100~TeV band are needed. Such an extension can hardly be done with space-based detectors, because the electron spectrum is very soft and relatively small area space-based detectors inevitably run out of statistics. The most promising possibility for extension of the measurements to higher energies is with the ground-based detectors.  Ground-based experiments detect electrons and positrons by observing Extensive Air Showers (EAS) that they initiate while penetrating into the atmosphere. The main challenge of the ground-based measurement is to distinguish the signal of electron  induced EAS  from  the background of EAS initiated cosmic ray protons and atomic nuclei. The electron flux is at the level of $\lesssim 10^{-4}$ of the nuclei flux at 10~TeV, see Fig. \ref{fig:electron_spectrum}.  Different dedicated background rejection techniques based on fitting of numerically predicted templates of "electron-like" EAS images to the observed EAS image data have been used to achieve rejection of proton and nuclei background  in the HESS data \cite{Aharonian:2008aa,kerszberg_icrc,kerszberg_thesis,kraus_thesis,kolitzis_thesis}.  The difficulty of this approach can be illustrated by comparison of the HESS measurement up to 20 TeV reported in conference proceedings \cite{kerszberg_icrc} and several PhD theses \cite{kerszberg_thesis,kraus_thesis,kolitzis_thesis}, with different types of background rejection techniques providing different results and indicate a large systematic uncertainty of the analysis.   

An alternative possibility for discrimination between the EAS induced by protons and atomic nuclei  and EAS induced by electrons and positrons is provided by  the measurement of muon content of the EAS. This technique is used by KASCADE \cite{Apel:2017ocm}, Carpet-2 \cite{Dzhappuev:2018qlt} and LHAASO \cite{Bai:2019khm} experiments to measure the identity of the primary cosmic ray  particles in the energy range above 100~TeV or above 1~PeV. Extension of the "muon tagging" approach for cosmic ray proton and nuclei background suppression toward lower energies is considered for the SWGO experiment \cite{Albert:2019afb}. It is however challenging because of decrease of density of muons on the ground associated to the decrease of the energy of the primary cosmic ray particles.  

Below we show that a deployment of a large continuous muon detector counting (rather than just sampling) muons at the detection level is necessary  to achieve efficiency of rejection of hadronic EAS background by a factor $>10^5$ needed for the measurement of \epm\ spectrum in 10-100~TeV energy range. We discuss possible implementations of such large muon detector. 

\section{Large water Cherenkov muon detector for 10-100 TeV EAS}

Muons are conventionally detected in the EAS experiments by particle detectors, like scintillator pads (e.g. in Carpet-2 experiment \cite{Dzhappuev:2018qlt}) or water tanks (like in LHAASO experiment  \cite{Bai:2019khm}) buried underground, or shielded by dense material layer (like in KASCADE experiment \cite{Apel:2017ocm}). Otherwise, muons can also be identified in surface Water Cherenkov Detector Arrays (WCDA) like those of HAWC \cite{Zuniga-Reyes:2017znm} and LHAASO because they produce much larger Cherenkov light signal with specific timing signature, due to the deep penetration. Muon identification in WCDA placed on the surface (rather than under-ground) is difficult because of the mixing of the muon Cherenkov light signal with the signal of electromagnetic EAS component. Finally, deep underwater Cherenkov detectors, like ANTARES \cite{Aslanides:1999vq}, Baikal-GVD \cite{Avrorin:2019dli} and Km3NET \cite{Bagley:2009wwa} are able to track trajectories of individual muons and measure their energy, to obtain information about  neutrinos that produce muons in interactions with material surrounding the detector. 

10-100~TeV EAS produced by protons generate only $10^2-10^4$ muons scattered across large area within $\sim 400$~m around the shower axis \cite{Antoni:2000mg,LAGUTIN2001274}, so that the muon density is as low as $\le 10^{-2}$~m$^2$. Low density of muons on the ground complicates their detection. A muon passing through water  with the speed faster than the speed of light in the medium generates some $Y_{Ch}\simeq 200 $~ph/cm Cherenkov photons in the 400-700~nm wavelength range. The number of photoelectrons detectable by a large (e.g. $d_{pmt}=10''=25$cm) photomultiplier tube (PMT)  with photon detection efficiency $\kappa$ placed at a distance $R$ from the particle track is 
$n_{p.e.}\simeq Y_{ch} d_{pmt}^2/ (8 R\tan(\theta_{Ch}))\simeq 5\left[\kappa/0.3\right]\left[d_{pmt}/25\mbox{ cm}\right]^2\left[R/10\mbox{ m}\right]^{-1}$
where $\theta_{Ch}\simeq 41^\circ$ is the Cherenkov angle in water. 
Adopting $n_{p.e.}\sim 5$ as a threshold for detection of particle track in water Cherenkov detector, we find that  single large photo-multiplier tube can detect muons passing at a distance as large as  $R\simeq 10\left[d_{pmt}/25\mbox{ cm}\right]^2$~m. Thus, deployment of optical sensors on a sparse grid with spacing smaller $\sim 2R$, in an light-tight  compartment at a depth range between $5$~m and $5$~m$+R$ (to suppress the electron/positron/gamma component of the EAS) can be sufficient. 

The 10-100~TeV EAS are able to reach the sea level and it is in principle not necessary to place the large muon detector at high altitude.  This allows to  directly use the optical modules of neutrino telescopes like  ANTARES \cite{Amram:2001mi} or Baikal-GVD \cite{Avrorin:2016por} and deploy the detector in the same water reservoir  as the original neutrino detector (sea or lake), partially re-using the available experimental infrastructure. Otherwise, the WCDA consisting of two-layer water tanks like foreseen for SWGO \cite{Albert:2019afb} can be considered.  

To determine the efficiency of the muon detector necessary for "muon tagging" of proton and nuclei induced EAS we have performed Monte-Carlo simulations of EAS  using CORSIKA package (\url{https://www.iap.kit.edu/corsika/index.php}, version 75600),  We have assumed two-layer detector setup at sea level altitude with the top layer counting particles with energies at least 100~MeV crossing the water surface and bottom layer sampling muons and hadrons with energies at least 1.5~GeV reaching the 5~m depth.   

The electron flux is by up to five orders lower than the overall cosmic ray flux in the 10-100~TeV energy range (see Fig. \ref{fig:electron_spectrum}).  Thus, the detector has to assure that no more than 1 in $10^5$ hadronic EAS is mis-interpreted as an electron / positron induced EAS.  To verify this, we have simulated $7\times 10^5$ proton-induced EAS in the broad energy range 1-100 TeV and $10^4$ electron-induced EAS, saving the particle content on the ground (hadrons and muons with energies above 300~MeV, electrons, positrons and gammas with energies above 30 MeV).  The EAS were allowed to impact anywhere within the extent of the detector.

\begin{figure}
\includegraphics[width=\columnwidth]{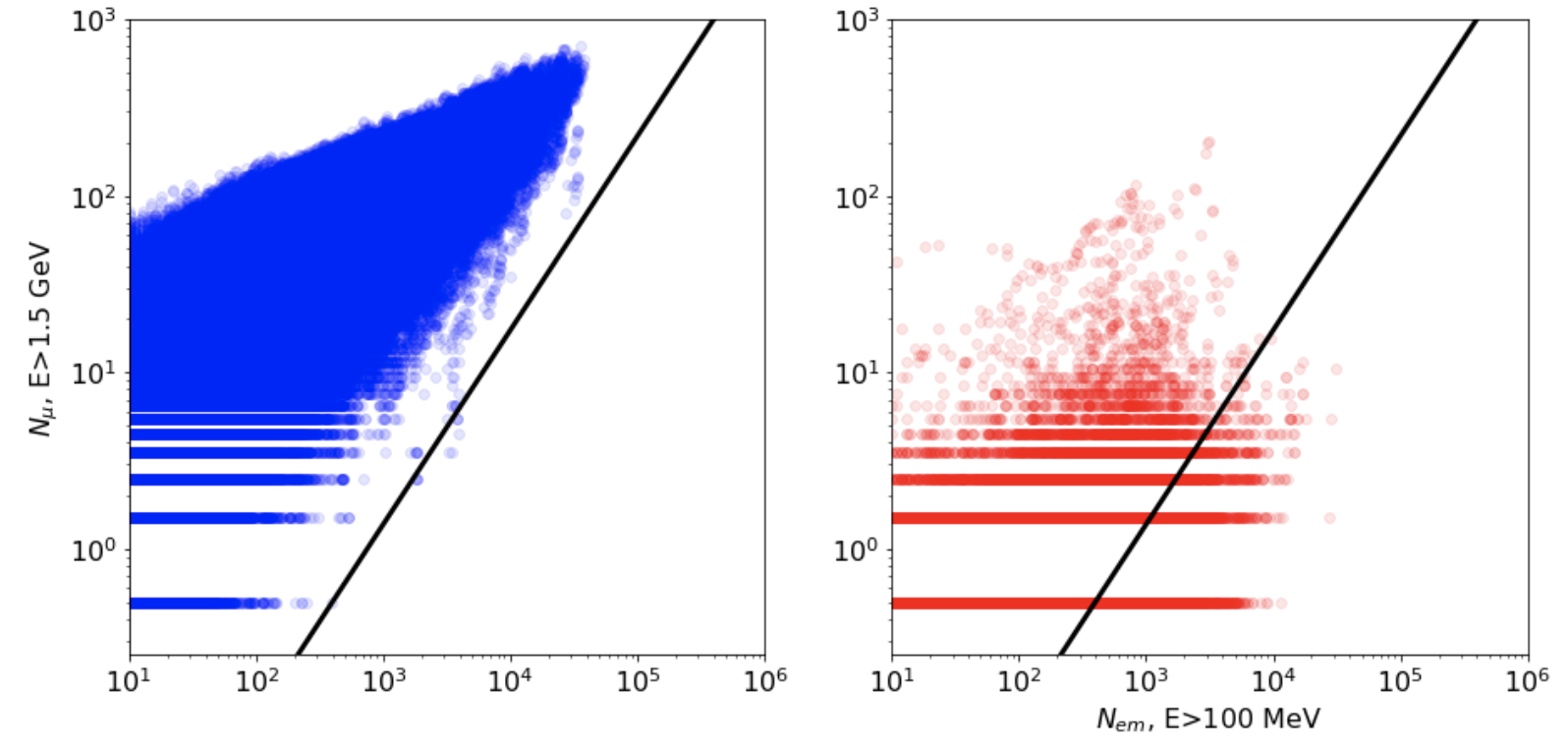}
\caption{Muon+hadron vs Electron-positron-gamma counts for the EAS impacting the $150\times 150$~m$^2$ area of \mwcda\   for proton induced EAS (left) and electrons (right). We have replaced zero muon counts with "0.5 count" value, to show the EAS without muon counts in this logarithmic plot.  Black line shows possible electron-hadron separation cut. }
\label{fig:Nmu_Cher}
\end{figure}

The statistics  of electrons, positrons and gammas with energies above 100~MeV, $N_{em}$ and statistics of muon and hadron counts with $E>1.5$~GeV for all EAS is  shown in Fig. \ref{fig:Nmu_Cher} for the selection of EAS falling within $150\times 150$~m$^2$ central part of a $300\times 300$~m$^2$ detector,  100\% efficient in counting both electromagnetic cascade particles and muons. 
To reject the hadronic EAS background, we have considered a cut shown by the straight line in  Fig. \ref{fig:Nmu_Cher} (cut optimisation has to be a subject of dedicated study). Only events to the right of the cut line or events with $N_\mu=0$ were accepted. 

\begin{figure}
\includegraphics[width=\columnwidth]{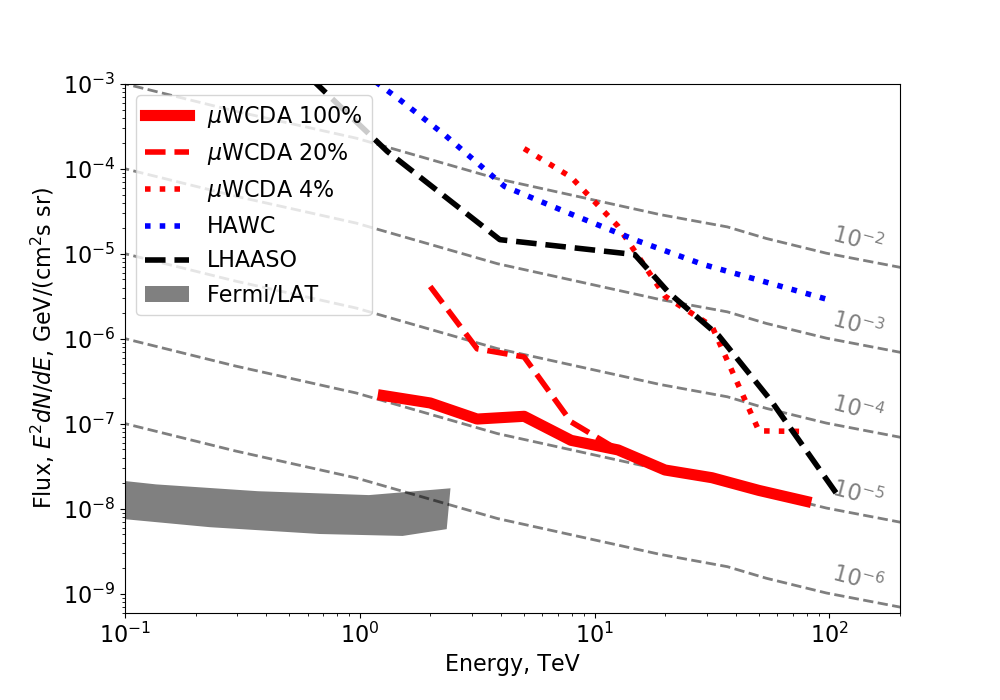}
\caption{Residual cosmic ray proton and nuclei background level for the setup with the muon detector layer underlying 100\% (red thick solid line), 20\% (dashed thin red line) and 5\% (red dotted line)  of WCDA. For comparison  we show the background levels in selected detectors: LHAASO (black thin solid line) and HAWC (blue thin solid line).}
\label{fig:bkg}
\end{figure}

If the muon layer of the detector is not continuous, only a fraction of the muons reaching the ground level would be counted. For example, LHAASO muon detector component is covers only about 4\% of the area. This reduces the statistics of the muon signal and hence the efficiency of electron-hadron shower discrimination. We have verified that our method of estimation of hadronic background rejection is consistent with the published estimated of backgorund rejection efficiency for LHAASO \cite{Bai:2019khm} if we assume that only 4\% of the muons are detectable. This is illustrated in Fig. \ref{fig:bkg}, where we compare the efficiency of rejection of proton shower background for different efficiencies of muon detection: 0.04 (LHAASO-type muon detector), 0.2 and 1 (continuous muon detector like that foreseen fro SWGO). 
One can see that deployment of continuous detector is necessary if one aims at backgorund rejection level $\sim 10^{-5}$
at 10 TeV. Decployment of muon detectors on a grid two times denser than that implemented in LHAASO would still not be sufficient. This result does not depend on the assumption about the altitude of the detector. In our simulations we counted muons reaching the sea level, but the number of muons does not change significantly with altitude in a wide altitude range, from 0 to about 4.5 km for the energy range of interest. We have verified this be repeating the simulations for the detector at 4.5 km altitude. 

\begin{figure}
\includegraphics[width=\columnwidth]{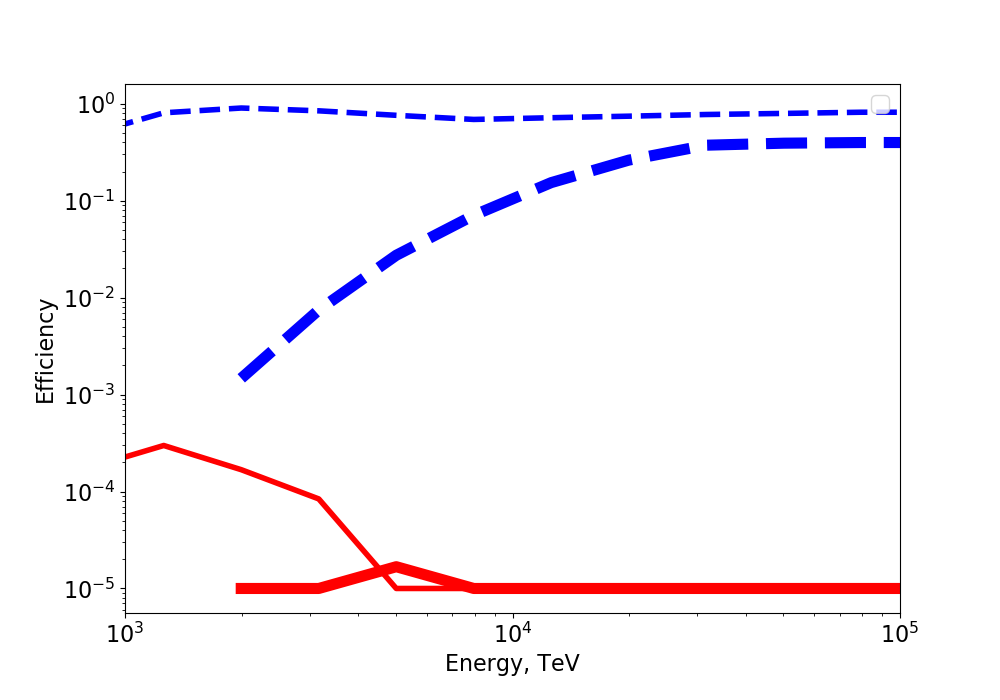}
\caption{Efficiency of the cut separating hadronic and \epm\ EAS as a function of energy, for electrons  (blue dashed lines) and protons (red solid lines) for a detector at sea level (thick lines) and at 4.5 km altitude (thin lines). }
\label{fig:efficiency}
\end{figure}

The altitude of the detector most strongly affects the statistics of the electromagnetic shower component. Decrease of the number of electromagnetic cascade particles at low altitude increases the energy threshold of the experiment. This is illustrated in Fig. \ref{fig:efficiency} in which the  thick blue dashed line shows the energy dependence of  efficiency of selection of electron EAS for the sea-level detector. This efficiency approaches $\sim 0.5$ in the energy range above 10~TeV.  The decrease of efficiency below 10~TeV is explained by the "quality" cut $N_{em}>100$ that we have imposed on the selection of both proton and electron events. We have imposed this cut to assure that the detected events can be properly analyzed and their energy  estimated with reasonable accuracy). Also this cut has to be optimized via a detailed study for specific detector setups. The same cut for a detector at 4.5 km altitude provides a nearly 100\% detection efficiency for electron showers down to TeV energy (thin dashed line). 

The HAWC and LHAASO WCDA have areas comparable to or larger than that of the muon continuous detector considered in our analysis. In principle, both WCDAs can detect all muons passing through them. They can also distinguish muon signal form the signal of electromagnetic EAS component, because single muons crossing the full depth of the water tanks produce larger Cherenkov signal. However, from Fig. \ref{fig:bkg} one can see that in spite of detecting similar amount of muons in 10-100~TeV EAS, HAWC and LHAASO WCDAs do not reach the efficiency of rejection of hadronic EAS comparable to that of the large muon detector. This is explained by the difficulty of identification of the muon signal on top of strong electromagnetic signal that dominates the EAS signal in WCDA not screened by the $5$~m water depth. Thus, an efficient $\mu$WCDA has to include a dedicated muon detection layer, rather than aim at muon detection on top of a dominant electromagnetic signal in single WCDA layer. 

\begin{figure}
\includegraphics[width=\columnwidth]{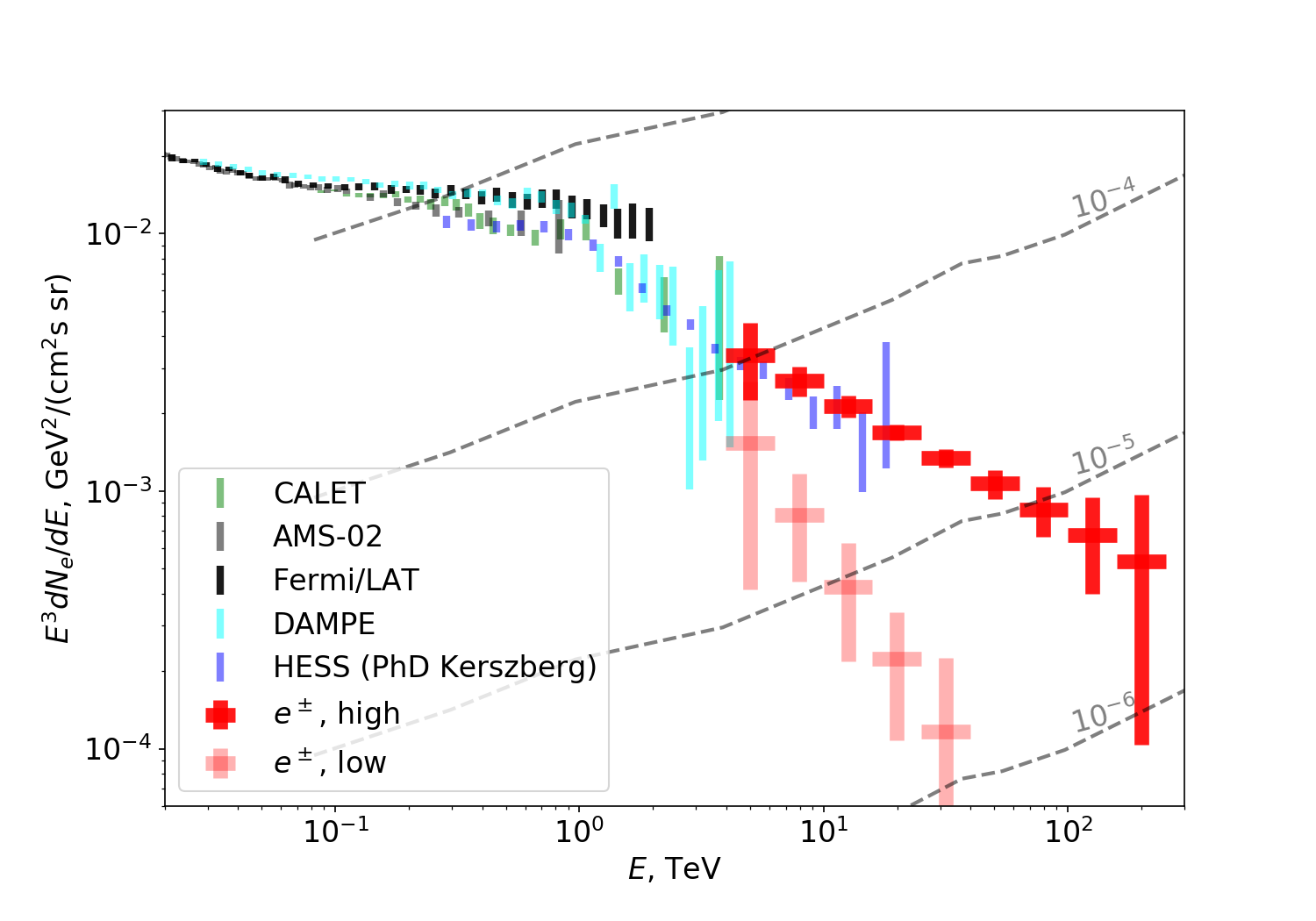}
\includegraphics[width=\columnwidth]{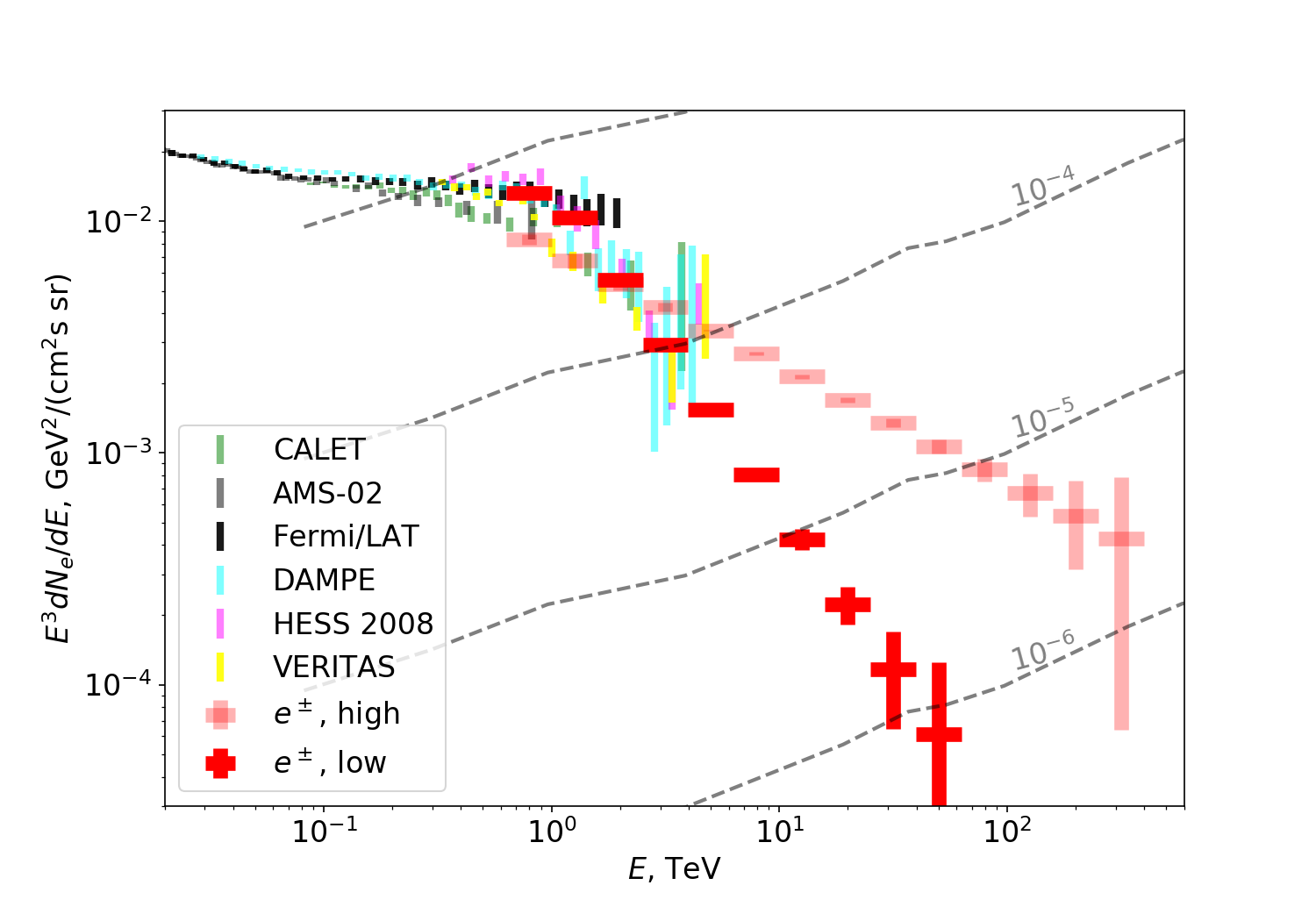}
\caption{Expected quality of electron spectrum measurement with WCDA detector with 100\% efficient muon detection layer at sea level altitude  (top), and at 4.5 km altitude (bottom).  Experimental data are the same as in Fig. \ref{fig:electron_spectrum}. Two possible high-energy powerlaw extrapolations are considered, with higher and lower flux levels with opaque and semi-transparent red color.  Full red color highlights the high flux (top) and low flux (bottom) extrapolations for respectively the sea level and high-altitude detectors.   }
\label{fig:electron_spectrum_measurement}
\end{figure}

Efficient suppression of hadronic background in WCDA with continuous muon detection layer enables measurement of electron spectrum in the energy range above 10~TeV. This is shown in Fig. \ref{fig:electron_spectrum_measurement}. The top panel of this figure shows numerically estimated quality of measurements of electron spectrum with the sea-level detector for possible "high" and "low" flux extrapolations of the electron spectrum measurements into 10-100 TeV range. To make these estimates we have assumed that the effective collection area for both proton background and electrons is the geometrical area of the detector times the efficiency of electron selection shown in Fig. \ref{fig:efficiency}. The angular acceptance of the WCDA is assumed to be $30^\circ$ around zenith, the exposure time is one year. The errorbars are determined by the statistical error for the electron signal detected on top of the residual hadronic background. A systematic error at the level of $5\%$ of the hadronic background is added. 

One can see that the flux at the level of the high flux powerlaw extrapolation from sub-10-TeV range is readily detectable up to 100~TeV and beyond on top of the residual hadronic background with an experimental setup using the $100\%$ efficient muon detection layer.  Remarkably, the electron spectrum can, in principle, be measured starting from the TeV energy range, in spite of the low efficiency of electron detection with an experimental setup at the sea level altitude. This is explained by the steepness of electron spectrum. The electron flux becomes as high as $\sim 10^{-3}$ of the cosmic ray flux at TeV. Even if the efficiency of electron detection is as low as $10^{-3}$ at this energy, suppression of the background of cosmic ray nuclei by a factor $\sim 10^{-5}$ still leaves the electron signal statistics at the level of about 10\% of the nuclei background statistics. 

If the electron spectrum above 1 TeV is as steep as the lowest flux allowed  by   VERITAS \cite{Archer:2018chh} or the 2008 HESS  \cite{Aharonian:2008aa} measurements (this would mean that the most recent HESS analysis \cite{kerszberg_icrc,kerszberg_thesis} still suffers from residual hadronic background contamination), the electron count statistics in a sea-level detector decreases below the residual cosmic ray nuclei statistics in all energy bins. This explains large errorbars of the measurements for this case (lower pale red data points in the top panel of Fig. \ref{fig:electron_spectrum_measurement}). In this case it might be essential to increase the efficiency of detection of electromagnetic component of the EAS, either by supplementing the sea-level detector with an IACT system or by deploying the detector at high altitude, as foreseen  in the SWGO design. The expected improvement of the quality of measurements with the high-altitude detector is shown in the bottom panel of Fig. \ref{fig:electron_spectrum} where the expected quality of measurements of electron spectrum with a two-layer WCDA at 4.5 km altitude is shown. The high-altitude and sea-level detectors provide similar quality of measurements above approximately 20~TeV. At lower energies, only the high altitude setup has sufficiently high efficiency of detection of electron induced EAS to provide high quality measurement.   One can see that in this case, the electron spectrum measurements can also have a good overlap with the measurements by space-based detectors.

\section{Discussion and conclusions}

We have shown that measurements of cosmic ray \epm\ spectrum up to 100~TeV are possible using an experimental  setup that includes an EAS array that provides nearly 100\% efficiency of muon detection across a $150\times 150$ m$^2$ area.  Such a detector can be implemented as an under-water Cherenkov detector combined with a conventional WCDA, as foreseen in the design of SWGO. Moreover, if the electron flux level at 10~TeV is as high as suggested by the most recent HESS data analysis,  such detector does not necessarily need to be located at high altitude. 

The most straightforward possibility for the sea-level setup is to implement the under-water muon detector re-using the development of water-Cherenkov neutrino detectors. The optical modules of neutrino detectors like ANTARES  \cite{Aslanides:1999vq}, Baikal-GVD \cite{Avrorin:2019dli} or Km3NET \cite{Bagley:2009wwa}
can be directly deployed at shallow depth in a two-storey grid (at depths about 5 and 10-15 m) within light-tight sections. Estimates of the Cherenkov light intensity show that the layout of the muon detection layer can be rather sparse, with module separation up to 20 m if large 12'' photo-multiplier tubes are used.   In the absence of the high-altitude constraint, the large  shallow depth water-Cherenkov detector array  can even be implemented directly at the sites of the underwater neutrino observatories, to simultaneously serve as a part of cosmic ray veto for the neutrino experiment (similar to the IceTop array at the site of IceCube experiment). 

\bibliography{Electron_spectrum}
\end{document}